\documentclass[conference]{IEEEtran}
\IEEEoverridecommandlockouts
\usepackage{cite}
\usepackage{amsmath,amssymb,amsfonts}
\usepackage{amsmath}
\usepackage{algorithmic}
\usepackage{graphicx}
\usepackage{textcomp}
\usepackage{xcolor}
\def\BibTeX{{\rm B\kern-.05em{\sc i\kern-.025em b}\kern-.08em
    T\kern-.1667em\lower.7ex\hbox{E}\kern-.125emX}}
\begin{document}

\title{Revealing Microscopic Objects in Fluorescence Live Imaging by Video-to-video Translation Based on A Spatial-temporal Generative Adversarial Network
}

\author{\IEEEauthorblockN{Yang Jiao, Mei Yang}
\IEEEauthorblockA{\textit{Department of Electrical and Computer Engineering} \\
\textit{University of Nevada, Las Vegas}\\
Las Vegas, US \\
yangjiaousa@gmail.com,mei.yang@unlv.edu}
\and
\IEEEauthorblockN{Mo Weng}
\IEEEauthorblockA{\textit{School of Life Sciences} \\
\textit{University of Nevada, Las Vegas}\\
Las Vegas, US \\
mo.weng@unlv.edu}
}

\maketitle

\begin{abstract}
In spite of being a valuable tool to simultaneously visualize multiple types of subcellular structures using spectrally distinct fluorescent labels, a standard fluoresce microscope is only able to identify a few microscopic objects; such a limit is largely imposed by the number of fluorescent labels available to the sample. In order to simultaneously visualize more objects, in this paper, we propose to use video-to-video translation that mimics the development process of microscopic objects. In essence, we use a microscopy video-to-video translation framework namely Spatial-temporal Generative Adversarial Network (STGAN) to reveal the spatial and temporal relationships between the microscopic objects, after which a microscopy video of one object can be translated to another object in a different domain. The experimental results confirm that the proposed STGAN is effective in microscopy video-to-video translation that mitigates the spectral conflicts caused by the limited fluorescent labels, allowing multiple microscopic objects be simultaneously visualized.
\end{abstract}


\begin{figure*}[]
\centering
\includegraphics[width=1.0\textwidth]{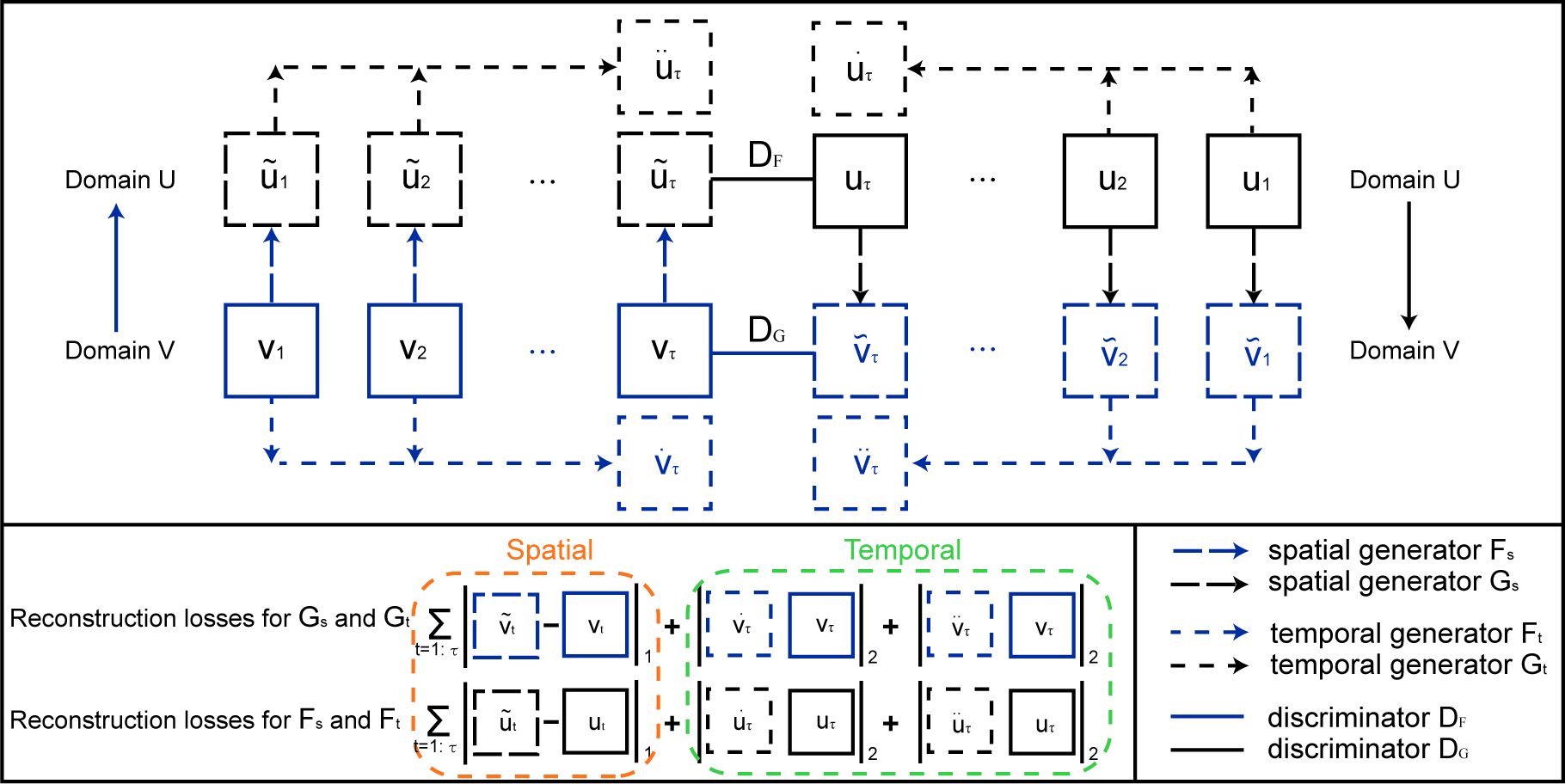}
\caption{STGAN framework. $\tilde{u_t}$ and $\tilde{v_t}$ are outputs of spatial generators. $\tilde{u_t}=F_s(v_t)$ and $\tilde{v_t}=G_s(u_t)$. $\dot{u_t}$ and $\dot{v_t}$ are outputs of temporal generators based on input frames. $\dot{u_t}=F_t(u_{1:\tau-1})$ and $\dot{v_t}=G_t(v_{1:\tau-1})$. $\ddot{u_\tau}$ and $\ddot{v_\tau}$ are outputs of temporal generators based on fake frames. $\ddot{u_\tau} = F_t(\hat{u_{1:\tau-1}})$ and $\ddot{v_\tau} = G_t(\hat{v_{1:\tau-1}})$
}
\label{fig:fig1}
\centering
\end{figure*}

\section{Introduction}
In fluorescence microscopy, samples labeled with fluorescent probes (a.k.a. fluorophores) are one of the most versatile optical imaging methods. A variety of fluorophores can be used to visualize biological events and live objects, including structures made of proteins, lipids, and ions \cite{giepmans2006fluorescent,sanderson2014fluorescence}. However, the wide emission spectra of most fluorophores limits the number of concurrent labels to 4-5 in a standard fluorescence microscope. In practice, the spectral cross-talk can be significant when using more than 3 fluorophores, which consequently limits the types of objects that can be observed and studied simultaneously. This is particularly problematic in live imaging of protein structures as it often relies on the use of fluorescent proteins. Compared with chemical dyes, there are limited choices of fluorescent proteins, and they are less photo-stable, difficult to apply, and have wider excitation and emission spectra. Microscopy video-to-video translation is a type of sample translation in computer vision that can be used to get around this limitation because successful translation produces the target object video to mimic its development process without actually occupying one fluorescent channel.
Sample translation in computer vision is a process that translates one possible representation of the sample (an image, avideo and etc.) to a different representation, and it has found its applications on microscopic image processing ranging from re-staining \cite{bayramoglu2017towards,jiao2022fine,jiao2023digitally,guo2024q,guo2024self}, transferring modalities \cite{han2017transferring,jiao2023learning,jiao2019deepquantify,guo2024learning,guo2024decorate3d,sun2023towards,sun2024revealing}, restoration \cite{quan2019removing,jiao2018automated,teng2017feasibility}, super-resolution \cite{wang2019deep,jiao2019multi,jiao2022modeling,luo2024and}, CT-to-MRI translation \cite{yi2019generative,zhang2018translating,zhou2023bandit}, to synthesis sample generation \cite{osokin2017gans,liu2025learning,chen2023improving,hairi2025enabling,Zhou2023}. Many of the studies on sample translation tend to use Generative Adversarial Network (GANs) \cite{goodfellow2014generative} and conditional GANs (cGANs) \cite{isola2017image}. cGANs are the general solution of sample translation that constrains the output space using the given conditions including class labels, text, images, etc. In the case of image conditioning, conditional GANs utilize structured losses to penalize the joint configuration of the output to control the structural difference between the output and target. 
For cGANS, an important property when used for sample translation is attributed to the uniqueness of the output-condition pair. Paired translation \cite{isola2017image,liu2016coupled,wang2018high,zhu2017toward,kutela2022determination,jiao2019self,gautam2022deep} assumes that a large amount of unique matching image pairs is accessible for model training. In practice, the challenge of pairing samples comes from studying the unpaired translations \cite{bansal2018recycle,chen2019mocycle,choi2018stargan,yi2017dualgan,jiao2024rethinking} that ignored unique matching image pairs. Specifically, when there exist domains U and V, and scenes A and B, paired translation approaches map domain U by conditioning on domain V, for the same scene. Unpaired translation approaches, on the other hand, tackle the translation of scene A in domain U by conditioning a different scene, scene B in domain V.
Note that, from the sample’s perspective, there exists subtle differences between image-to-image translation \cite{isola2017image,liu2016coupled,wang2018high,zhu2017toward,choi2018stargan,yi2017dualgan} and video-to-video \cite{bansal2018recycle,chen2019mocycle,zhou2022bandit,naughton2021elastica,sun2024comusion} translation. The image-to-image translation differs from the video-to-video translation in terms of the richness of the conditioning information. The image-to-image translation translates an input image to a different domain by conditioning the input in the spatial domain. It is applied to cases like translating between a semantic map and a realistic scene, or between artworks and realistic scenes. Other than image-to-image translation, video-to-video translation takes advantage on both spatial and temporal information which provides more constrains to optimize the transformation from one domain to another. For example, Recycle-GAN \cite{bansal2018recycle,guo2020deep,jiarui2025system,sun2024mooss} introduces temporal information by introducing recurrent temporal predictors to the translation. Mocycle-GAN \cite{chen2019mocycle,shang2024transitivity,guo2019pipeline,zhou2022bandit,Zhang2024} considers motion predictors in cross-domain translation to emphasize the motion information.
In this paper, the problem of microscopic object translation using fluorescence live images seeks for the spatial distribution of object (or domain) U/V based on the spatial and temporal distribution of object (or domain) V/U. Thus, it can be seen as a problem of paired video-to-video translation, or paired cross-domain video-to-video translation because the translations between U and V are valid independently and concurrently. Nevertheless, different from other cross-domain translation problems \cite{wang2019deep,yi2019generative,zhu2017toward,choi2018stargan,hong2020effect,hong2018cross}, the problem of microscopic object translation has its own unique challenges, including 
\begin{itemize}
\item The existence of a correlation between objects may be unknown. The translation may fail if U and V are independent. 
\item The strength of the correlation between objects varies. For example, the correlation is strong if the development of V only depends on U, or weak if the development of V depends on multiple different objects including U. 
\item The time shift and impact duration of correlation between objects vary. The impact of one object to another may not be real-time, and the impact may last for various durations. 
\item The morphology forms and spatial distributions of objects may differ tremendously.
In this paper, we first formulate the microscopy video-to-video translation problem. We then present a cross-domain video-to-video translation network, namely Spatial-temporal Generative Adversarial Network (STGAN), to learn the spatial and temporal relationship between microscopic objects and effectively translate microscopic objects. Experimental results of the proposed STGAN are presented and compared with existing approaches.
\end{itemize}

\section{Method}

\subsection{Problem definition of cross-domain translation}

Given two sets of sequential frames of spatial distributions from heterogeneous microscopic objects $U$ and $V$, $\{u_t\in U^T\}$,$\{v_t\in V^T\}$, suppose we know there exists a causal relationship between $U$ and $V$ with causal duration $\tau$ and time shift $s$.
Since the problem of cross-domain video-to-video translation by conditional GANs can be formulated as $p(\hat{V^T}|U^T)$.

By the Markov assumption, we factorize the conditional distribution into a production form:
\begin{equation}
 p(\hat{V^T}|U^T) = \Pi_{t}^{T}p(\hat{v_t}|\{u_{t-s},u_{t-s-1},...,u_{t-s-\tau}\})
\label{eq:eq1}
\end{equation}
\begin{equation}
 p(\hat{U^T}|V^T) = \Pi_{t}^{T}p(\hat{u_t}|\{v_{t+s},v_{t+s-1},...,v_{t+s-\tau}\})
\label{eq:eq2}
\end{equation}
where $\{u_{t-s},u_{t-s-1},...,u_{t-s-\tau}\}$ and $\{v_{t+s},v_{t+s-1},…,v_{t+s-\tau}\}$ are the $\tau$ corresponding frames of two microscopic objects $U$ and $V$, respectively. The objective of the problem of microscopy video-to-video translation is to maximize the probability of translation $p(\hat{V^T}|U^T)$ and $p(\hat{U^T}|V^T)$. For each frame, the probability of prediction by translation framework $\theta$ is
\begin{equation}
\begin{split}
& p(\hat{v_t}|\{u_{(t-s)},u_{(t-s-1)},...,u_{(t-s-\tau)}\}) = \\
&p(\theta(\{u_{t-s},u_{t-s-1},...,u_{t-s-\tau}\})|\{u_{t-s},u_{t-s-1},...,u_{t-s-\tau}\})
\label{eq:eq3}
\end{split}
\end{equation}

\begin{equation}
\begin{split}
& p(\hat{u_t}|\{v_{(t+s)},v_{(t+s-1)},...,v_{(t+s-\tau)}\}) = \\
&p(\theta(\{v_{t+s},v_{t+s-1},...,v_{t+s-\tau}\})|\{v_{t+s},v_{t+s-1},...,v_{t+s-\tau}\})
\label{eq:eq4}
\end{split}
\end{equation}
Therefore, to maximize $p(\hat{V^T}|U^T)$ and $p(\hat{U^T}|V^T)$, we need to improve the predictions by improving the translation framework $\theta$. In the next section, we will propose a translation framework to present an efficient $\theta$.

\subsection{Framework}

Without loss of generality, we assume that the casual duration between $U$ and $V$ is $\tau$ ($\tau\leq0$) and the time shift $s$ equals to zero ($s$ can be change to 0 by preprocessing). The proposed Spatial-temporal Generative Adversarial Network (STGAN) framework of cross-domain video-to-video translation for microscopic objects consists of two spatial generators ($G_s$ and $F_s$), two temporal generators ($G_t$ and $F_t$) and two discriminators ($D_G$ and $D_F$), as shown in Fig.  1. $G_s$ takes one frame of microscopy video of object $U$ as an input and predicts the spatial distribution of $V$, while $F_s$ takes one frame of $V$ and predicts the spatial distribution of $U$. $G_t$ and $F_t$ take $\tau-1$ frames from domain $U$ and $V$ to predict the next frame in domain $V$ and $U$, respectively. Unlike $Gs$ and $F_s$ that learn the spatial relationship between $U$ and $V$, $G_t$ and $F_t$ learn the temporal relationship between the frames of $U$ and $V$, respectively. $D_G$ and $D_F$ are discriminators that make judgments on real and fake frames in domain $U$ and $V$, respectively.

\begin{figure*}[]
\centering
\includegraphics[width=1.0\textwidth]{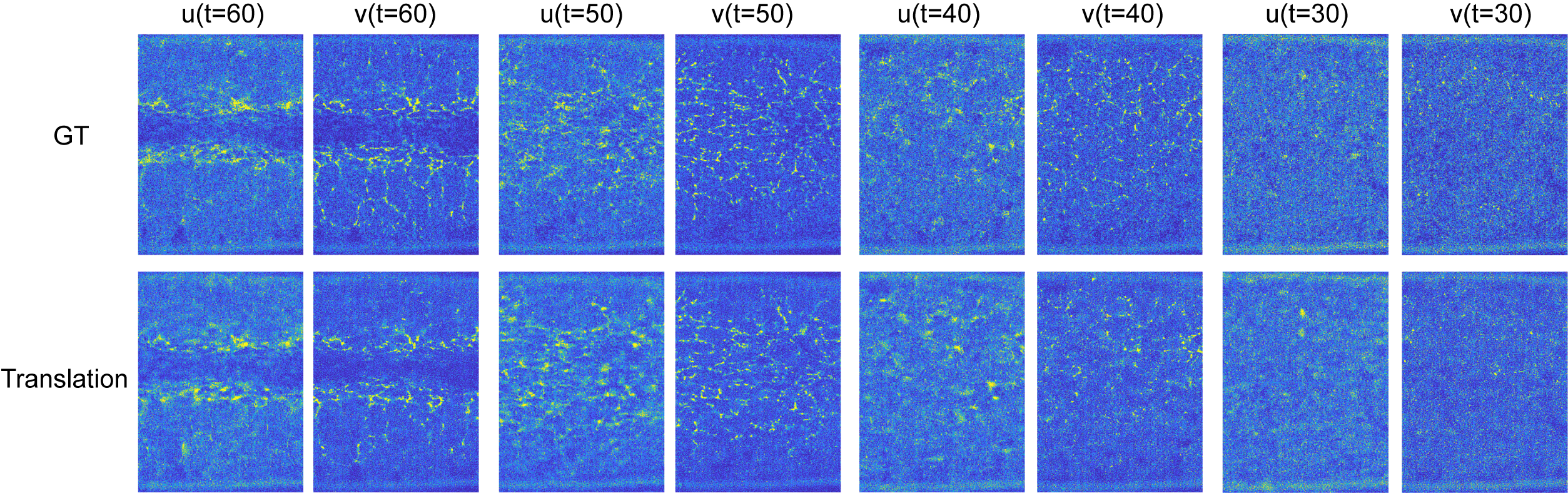}
\caption{Cross-domain translation results of STGAN. $u(t)$ is a frame of Myosin live image. $v(t)$ is a frame of Ajuba live image. $t$ is the time stamp of the live image.}
\label{fig:fig2}
\centering
\end{figure*}

\subsection{Objectives}
\subsubsection{Adversarial losses}
In our proposed framework,the generators $G_s$ and $F_s$ are trained with two independent adversatial losses: $L_{adv G_s}$ and $L_{adv F_s}$.
\begin{equation}
L_{adv G_s} = \Sigma_{t=1}^{\tau}l_{D_G}(G_s,D_G)
\label{eq:eq5}
\end{equation}
\begin{equation}
L_{adv F_s} = \Sigma_{t=1}^{\tau}l_{D_F}(F_s,D_F)
\label{eq:eq6}
\end{equation}

$l_{D_G}$ and $l_{D_F}$ are explained in Eqn. (7) and (8). let $u_t$ and $v_t$ denote the frames of $U$ and $V$ at time $t$, $\tilde{u_t}$ and $\tilde{v_t}$ are the fake frames at time $t$. $\tilde{v_t}$ and $v_t$ are the inputs to discriminator $D_G$, and $\tilde{u_t}$ and $u_t$ are the inputs to discriminator $D_F$. The loss functions of $D_G$ and $D_F$ are defined as 
\begin{equation}
l_{D_G}(G_s,D_G) = log(D_G(v_t)) + log(1-D_G(\tilde{v_t}))
\label{eq:eq14}
\end{equation}
\begin{equation}
l_{D_F}(F_s,D_F) = log(D_F(u_t)) + log(1-D_F(\tilde{u_t}))
\label{eq:eq15}
\end{equation}

\subsubsection{Spatial reconstruction losses}
The previous works \cite{zhao2016loss} have found it beneficial to mix the adversarial loss with reconstruction losses, and discussed the various reconstruction losses including L1, L2, and SSIM \cite{wang2004image,guo2022data,luo2021functional,jiarui2023dynamic,sun2022dynamic} loss. Unlike L2 loss, L1 loss is a widely accepted loss in image and video translation because it encourages less blurring. 
\begin{equation}
l_{l_1}(G_s) = \Sigma_{t=1}^\tau||\hat{v_t},v_t||_1
\label{eq:eq7}
\end{equation}
\begin{equation}
l_{l_1}(F_s) = \Sigma_{t=1}^\tau||\hat{u_t},u_t||_1
\label{eq:eq8}
\end{equation}

\subsubsection{Temporal reconstruction losses}
To train the temporal generators, we first consider the loss of next frame prediction. Taking the real frames as input, the temporal loss is the error between the predicted frame and the real frame.
\begin{equation}
l_{t}(G_t) = ||\dot{v_\tau},v_\tau||_2 = ||G_t(v_{1:\tau-1}),v_\tau||_2
\label{eq:eq9}
\end{equation}
\begin{equation}
l_{t}(F_t) = ||\dot{u_\tau},u_\tau||_2 = ||F_t(u_{1:\tau-1}),u_\tau||_2
\label{eq:eq10}
\end{equation}

Next, we apply the temporal generators on the fake frames, to further compose the temporal and spatial losses.
\begin{equation}
l_{ts}(G_t,G_s) = ||\ddot{v_\tau},v_\tau||_2 = ||G_t(\hat{v_{1:\tau-1}}),v_\tau||_2
\label{eq:eq11}
\end{equation}
\begin{equation}
l_{ts}(F_t,F_s) = ||\ddot{u_\tau},u_\tau||_2 = ||G_t(\hat{u_{1:\tau-1}}),u_\tau||_2
\label{eq:eq12}
\end{equation}

\subsubsection{Full objectives}
For the cross-domain translation in our proposed STGAN framework, the final objective is achieved by optimizing the adversarial, the spatial reconstruction losses, and the temporal reconstruction losses.
\begin{equation}
\begin{split}
G_s,F_s,G_t,F_t &= argmin(L_{adv G_s} + L_{adv F_s} + \\
&\lambda_sl_{l_1}(G_s) + \lambda_sl_{l_1}(F_s) + \lambda_tl_{t}(G_t) + \\
&\lambda_tl_{t}(F_t) + \lambda_tl_{ts}(G_t,G_s) + \lambda_tl_{ts}(F_t,F_s))
\label{eq:eq13}
\end{split}
\end{equation}

\subsubsection{STGAN outputs}
From above, we know that the proposed STGAN has six types of output for cross-domain translation: $\tilde{u}, \dot{u}, \ddot{u}, \tilde{v}, \dot{v}, \ddot{v}$. For translation from domain $V$ to $U$, we either take $\tilde{u}$ as the output, $\hat{u}=\tilde{u}$ or we average the spatial and temporal translations to further smooth the output, $\hat{u}=(\tilde{u}+\ddot{u})/2$. Translation from $U$ to $V$ can take either $\hat{v}=\tilde{v}$ or $\hat{v}=(\tilde{v}+\ddot{v})/2$ as the output.

\section{Experiment}
\begin{figure}[h]
\includegraphics[width=0.45\textwidth]{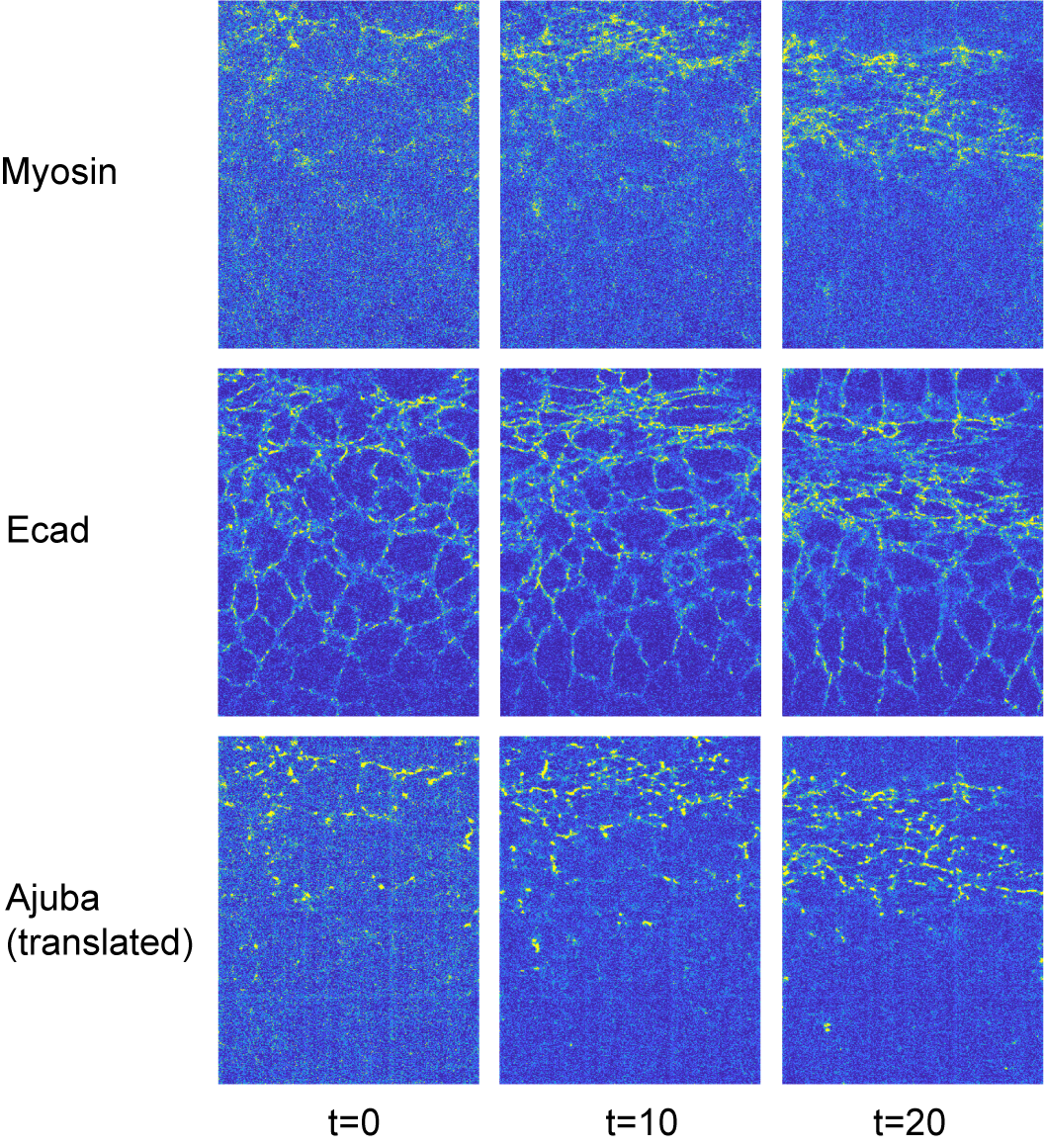}
\caption{Cross-domain translation evaluation.}
\label{fig:fig3}
\end{figure}

\textbf{Data description}. In our data sets, Myosin and Ajuba proteins are genetically fused with fluorescent proteins mCherry and GFP respectively, to enable the simultaneous live imaging of the two proteins in Drosophila embryos. The embryos were mounted for imaging at early developmental stages and continued to develop while being imaged. Myosin and Ajuba do not spatially overlap extensively, but there seems to be some spatial and temporal correlation, consistent with the Myosin-dependent recruitment of Ajuba. Signals emitted from Myosin-mCherry and Ajuba-GFP correspond to domain U and V, respectively. We collect the 60 frames of live imaging from two embryos, and crop the frames into 128*128 for training/validation (4k/1k).

\textbf{Implementation}. Considering the fairness of evaluation, all evaluated frameworks adopt the generator and discriminator networks from Pix2pix \cite{isola2017image}. The spatial generators are U-Nets that take one frame as input. The temporal generators are U-Nets that concatenate multiple frames as input. The discriminators are 70x70 PatchGAN \cite{isola2017image}. The weights of spatial and temporal reconstruction objectives $\lambda_s$ and $\lambda_t$ are set as $10^2$ according to \cite{isola2017image} and 10, respectively. We used an Adam optimizer \cite{diederik2014adam} with a batch size of 8. The networks were trained with learning rate $\iota=2*10^{-4}$. 

\textbf{Experiments}. We employ three popular translation networks as baselines, including Pix2pix, CycleGAN and RecycleGAN that cover paired and unpaired image-to-image and video-to-video translation. MSE, SSIM, and PSNR are used as evaluation metrics.

\section{Results}
\textbf{Efficiency of temporal information in microscopic object translation.} Table 1 shows the MSE, SSIM, and PSNR evaluation results of STGAN and three networks in comparison. Compared with Pix2pix and CycleGAN, RecycleGAN and the proposed STGAN present higher evaluation scores, which proves the effect of using temporal information in microscopic object translation. Furthermore, in translating U to V and V to U, STGAN obtains the best results regarding all evaluation metrics. The evaluation results confirm that STGAN is advantageous in microscopic object translation. Fig. 2 presents the translation results of STGAN compared with the ground truth.

\textbf{Challenges of microscopy video-to-video translation.} As mentioned in the introduction, correlations between different microscopic objects and different time points may vary tremendously. Noticeably, in Table 1 and Fig. 2, the error of STGAN from U to V is smaller than that from V to U. Our hypothesis is that the correlations between U and V are unequal bi-directionally. In addition, the translation using STGAN is more successful at later time points, which may be because the correlation is strengthened at later time points.

\textbf{Simulating protein live imaging.} Consider the limit of labeled objects in fluorescence microscopy imaging, using the proposed STGAN framework, we can predict the spatial distribution of an object by another. Therefore, the limitation of fluorescence absorption spectra can be overcome and the spatial distribution of multiple proteins can be obtained. Fig. 3 shows two live images of Myosin and Ecad which are simultaneously captured by two different fluorescence probes. Via the proposed STGAN, the video of Ajuba is obtained by translating the video of Myosin to realize the simultaneous observation of three proteins.

\section{Conclusion}
In fluorescence microscopy, the spectral cross-talk between fluorophores may limit the visualization of up to three types of microscopic objects simultaneously, which restricts research that studies multiple objects in an individual sample. In this paper, we propose to overcome this limit by microscopy video-to-video translation because successful translation produces the target object video to mimic its development process without occupying one fluorescent channel. To achieve an efficient microscopy video-to-video translation, we propose the translation framework STGAN which reveals both spatial and temporal relationships between microscopic objects. Our experiment confirms that STGAN effectively predicts the spatial distribution of a microscopic object over time by conditioning on the video of another microscopic object, and generates superior translation results compared to baseline translation frameworks.

\bibliographystyle{splncs04}
\bibliography{egbib}

\end{document}